\documentclass[pdflatex]{sn-jnl}

\usepackage{graphicx}%
\usepackage[numbers]{natbib}
\usepackage{lineno}
\usepackage{multirow}%
\usepackage{amsmath,amssymb,amsfonts}%
\usepackage{amsthm}%
\usepackage{mathrsfs}%
\usepackage[title]{appendix}%
\usepackage{xcolor}%
\usepackage{textcomp}%
\usepackage{manyfoot}%
\usepackage{booktabs}%
\usepackage{algorithm}%
\usepackage{algorithmicx}%
\usepackage{algpseudocode}%
\usepackage{listings}%
\usepackage[justification=centering]{caption}%
\usepackage{bm} 
\usepackage{booktabs}             
\usepackage{siunitx}              
\usepackage{diagbox}   
\usepackage{array}     
\theoremstyle{thmstyleone}%
\theoremstyle{thmstyletwo}%

\theoremstyle{thmstylethree}%

\begin{document}

\title[Article Title]{Mixing plant for JUNO liquid scintillator: Design, construction, installation and commissioning}


\author[1,2,3]{\fnm{Junyu} \sur{Shao}}

\author[1,2,3]{\fnm{Wenjie} \sur{Li}}

\author*[1,3]{\fnm{Xilei} \sur{Sun}}\email{sunxl@ihep.ac.cn}
\author[1,2,3]{\fnm{Tao} \sur{Hu}}
\author[1,3]{\fnm{Li} \sur{Zhou}}
\author[1]{\fnm{Yayun} \sur{Ding}}
\author[1,3]{\fnm{Xiao} \sur{Cai}}
\author[1,2,3]{\fnm{Boxiang} \sur{Yu}}
\author[1,3]{\fnm{Jian} \sur{Fang}}
\author[1,3]{\fnm{Yuguang} \sur{Xie}}
\author[1,3]{\fnm{Jie} \sur{Zhao}}
\author[1,3]{\fnm{Lijun} \sur{Sun}}
\author[1,3]{\fnm{Wanjin} \sur{Liu}}
\author[1,3]{\fnm{Hansheng} \sur{Sun}}
\author[1]{\fnm{Mengchao} \sur{Liu}}
\author[1,3]{\fnm{Xin} \sur{Ling}}

\affil*[1]{\orgname{Institute of High Energy Physics, Chinese Academy of Sciences}, \orgaddress{\city{Beijing}, \postcode{100049}, \country{China}}}

\affil[2]{\orgname{University of Chinese Academy of Sciences}, \orgaddress{\city{Beijing}, \postcode{100049}, \country{China}}}

\affil[3]{\orgname{State Key Laboratory of Particle Detection and Electronics}, \orgaddress{\city{Beijing}, \postcode{100049}, \country{China}}}

\abstract{The most challenging part of building the Jiangmen Underground Neutrino Observatory (JUNO) is the production of 20 kilotons of ultra pure Liquid Scintillator (LS). This paper presents the design, construction, installation, and commissioning of the LS Mixing Plant, a core facility dedicated to blending the primary organic solvent (LAB) with essential functional solutes (PPO, bis-MSB, and BHT). The main purpose of the Mixing Plant is to prepare and purify the concentrated Master Solution (MS) to achieve a low radioactive contamination background. The amount of radioactive contaminants in the MS are lowered by approximately two orders of magnitude after acid and water extraction, followed by a multi-stage filtration procedure. The purified MS is mixed with LAB and then diluted into the LS for JUNO experiments. Commissioning results of the LS verify that the Mixing Plant achieved its design goal, delivering ultra pure LS that satisfies the stringent radiopurity requirements for neutrino physics.}


\keywords{JUNO, Mixing Plant, Master Solution, liquid scintillator, radioactive contaminants}



\maketitle

\section{Introduction}\label{sec1}




The Jiangmen Underground Neutrino Observatory (JUNO), located 700 meters underground in Guangdong Province, China, is an international neutrino detection experiment with over 700 collaborators from 74 institutions. The experiment’s primary physics goals-determining the neutrino mass ordering and measuring oscillation parameters with sub-percent precision-demand a detector of unprecedented scale and sensitivity \cite{JUNO_PPNP}.

The Central Detector (CD), a 35.4-meter-diameter acrylic sphere, houses 20 kilotons of ultrapure liquid scintillator (LS) with an attenuation length exceeding 20 m \cite{JUNO:2015sjr}. This massive target volume is monitored by an integrated array of 17,612 20-inch and 25,600 3-inch photomultiplier tubes (PMTs), achieving a high optical coverage of 77.9\% \cite{He_2023}.

The JUNO LS consists of four organic components: linear alkylbenzene (LAB) as the solvent; 2,5-diphenyloxazole (PPO, 2.5 g/L) as the primary solute; 1,4-bis(2-methylstyryl)benzene (bis-MSB, 3 mg/L) as a wavelength shifter; and 2,6-di-tert-butyl-4-methylphenol (BHT, 42.8 mg/L) as an antioxidant to prevent aging of the LS. The production of the 20 kton of LS constitutes one of the major challenges of this experiment, primarily due to the stringent requirements regarding the radioactive contamination levels within the LS (as detailed in Table~\ref{tab1}).

\newcommand{\nuclide}[2]{\(\boldsymbol{^{#1}\mathrm{#2}}\)}
\begin{table}[h]
\centering
\caption{Radiopurity Requirements of LS for Neutrino Detection Channels (units: g/g) \cite{JUNO:2021kxb}}
\label{tab1}
\renewcommand{\arraystretch}{1.4} 
\begin{tabular}{|c|c|c|c|c|}
\hline
\textbf{Channels}&
\nuclide{238}{U}{} &
\nuclide{232}{Th}{} &
\nuclide{40}{K}{} &
\nuclide{210}{Pb}{} \\
\hline
\textbf{reactor neutrino}&
\SI{1e-15}{} &
\SI{1e-15}{} &
\SI{1e-16}{} &
\SI{1e-22}{} \\
\hline
\textbf{solar neutrino}&
\SI{1e-17}{} &
\SI{1e-17}{}&
\SI{1e-18}{} &
\SI{1e-24}{} \\
\hline
\end{tabular}
\end{table}



\begin{figure}[h]
\centering
\includegraphics[width=0.9\textwidth]{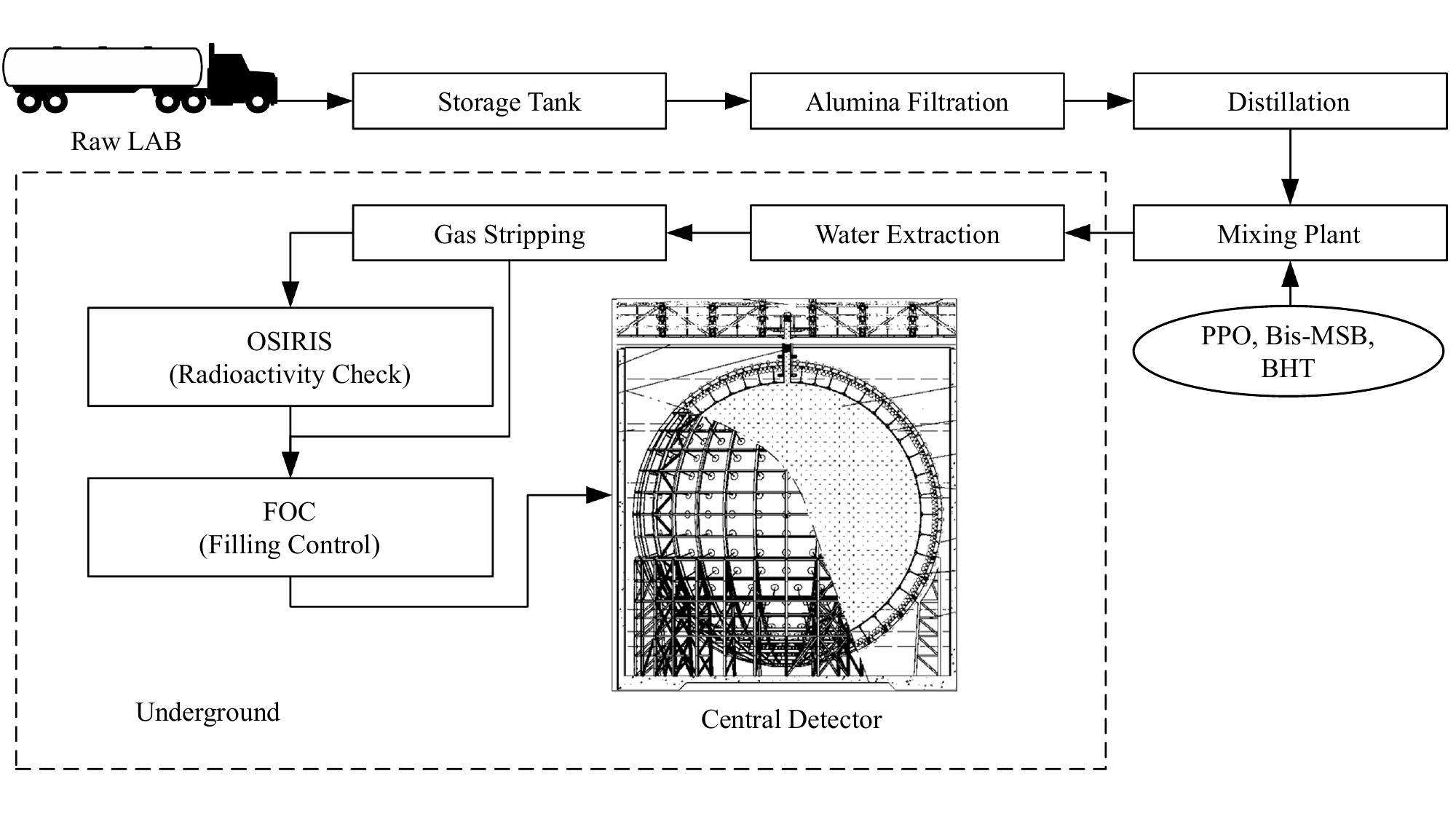}
\caption{LS purification flowchart. A comprehensive schematic detailing the multi-stage flow from raw LAB storage through alumina filtration and distillation, into the Mixing Plant for solute integration, followed by water extraction and gas stripping underground before final delivery to the Central Detector.}\label{fig1}
\end{figure}

However, the radioactivity levels of the four LS components supplied by commercial manufacturers did not meet the stringent radioactive background requirements of the experiment. Consequently, a multistage on-site purification process was essential. To achieve this, the JUNO LS team designed and constructed a dedicated ultrapure LS production line, comprising multiple purification modules along with an integrated quality control system, as shown in Fig.~\ref{fig1} \cite{ZHU2023167890}.

The sequence of subsystems begins with a \SI{5000}{\cubic\meter} storage tank, which stores the raw LAB and supplies it to the Alumina Filtration Plant (AFP) via long-distance pipelines. Serving as the first-stage purification facility, the AFP is dedicated to improving the optical clarity of the LAB \cite{ZHU2023167890}. Following the AFP, the Distillation Plant is employed to distill the LAB, effectively removing heavy metal radioisotopes such as uranium (U) and thorium (Th) \cite{LANDINI2024169887}.




The Mixing Plant facilitates the dissolution of PPO, bis-MSB, and BHT into LAB to prepare a concentrated Master Solution (M.S.), featuring a solute concentration 37 times higher than that of the standard-concentration LS. This M.S. is subsequently purified via liquid-liquid extraction using dilute nitric acid and high-purity water, followed by dual-stage filtration. The purified M.S. is subsequently diluted with LAB to yield the standard-concentration LS. Through these procedures, the concentrations of $^{238}\text{U}$ and $^{232}\text{Th}$ in the LS are reduced to below $10^{-16}$g/g.

The resulting LS is transported to a buffer tank located in the underground LS hall via an inclined-shaft pipeline at a flow rate of 7,000 L/h. Within this underground facility, several additional purification systems are deployed. These include the Water Extraction Plant, which extracts potassium (K), lead (Pb), radium (Ra), and other water-soluble contaminants from the LS \cite{YE2022166251}, and the Gas Stripping Plant, which eliminates dissolved gases such as radon (Rn), argon (Ar), and krypton (Kr) \cite{LANDINI2024169887}.

The radiopurity of the LS is continuously monitored by the Online Scintillator Internal Radioactivity Investigation System (OSIRIS). The injection of the LS into the Central Detector and its circulation within the online loop are managed by the Filling, Overflow, and Circulation (FOC) Plant. Furthermore, the high-purity water and nitrogen requisite for the LS purification processes are supplied by two auxiliary facilities: the High-Purity Water (HPW) Plant and the High-Purity Nitrogen (HPN) Plant \cite{LING2024111305}.


The development of the Mixing Plant spanned an eight-year period, commencing with preliminary design and proceeding through laboratory-scale process studies, pilot-scale validation, engineering design finalization, construction, installation, and commissioning. Results from the joint commissioning phase demonstrate that the facility strictly meets its design specifications and is fully capable of producing ultrapure LS. This paper provides a comprehensive overview of the LS mixing process, with a particular emphasis on the engineering, system integration, and operational validation of the Mixing Plant.

\section{Process Flow of LS Mixing}\label{sec2}



\begin{figure}[h]
\centering
\includegraphics[width=0.9\textwidth]{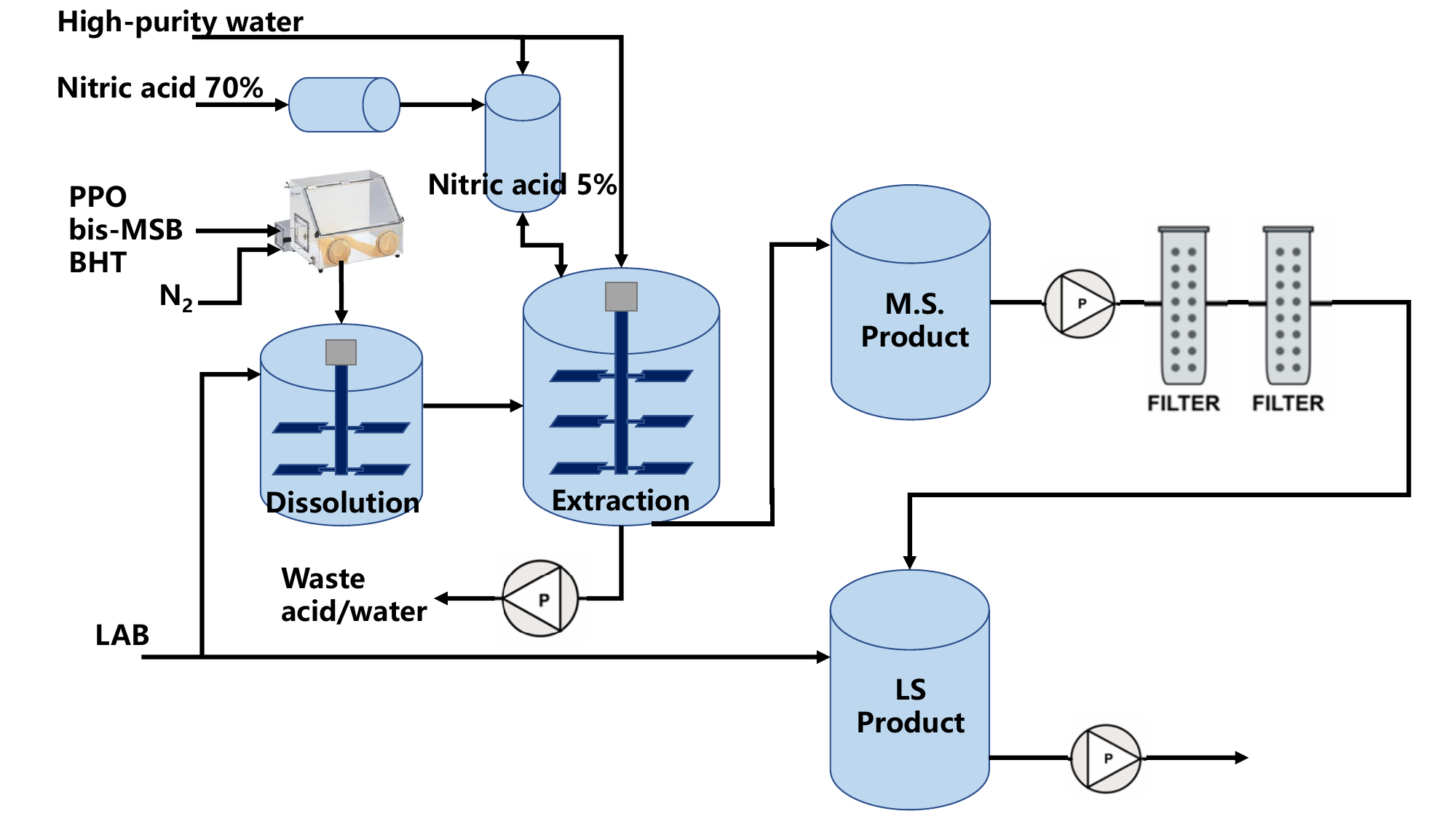}
\caption{LS mixing process. A detailed schematic illustrating the batch cycle of the Master Solution. This diagram depicts the weighing and feeding of PPO, bis-MSB, and BHT into the dissolution tank, the subsequent interaction between the organic and aqueous phases during nitric acid extraction, and the final two-stage filtration and dilution steps required prior to underground transport.}\label{fig2}
\end{figure}

As illustrated in Fig.~\ref{fig2}, the LS mixing process is primarily composed of solute feeding, dissolution, acid and water extraction, filtration, dilution, and final transfer. Given the target concentrations for PPO, bis-MSB, and BHT, alongside a continuous production flow rate of \SI{7000}{\liter/\hour}, the required daily feed amounts are calculated to be \SI{420}{\kilo\gram} of PPO, \SI{504}{\gram} of bis-MSB, and \SI{7.2}{\kilo\gram} of BHT.

Prior to use, each batch of solutes is precisely weighed and documented. To maintain radiopurity, all solutes are stored in vacuum-sealed packaging. The outer vacuum layer is removed inside a Class 100,000 cleanroom immediately before loading. The solutes are subsequently introduced into the dissolution tank via a nitrogen-purged glovebox, where they are dissolved to form the M.S..

The parameters for the M.S. concentration and dissolution temperature are optimized based on both solute solubility and purification efficiency. An excessively high concentration may lead to solute precipitation, whereas an overly dilute concentration can compromise purification performance. Each batch of M.S. is prepared by dissolving the solutes in 4200 L of LAB at a temperature of \SI{40}{\celsius}. The resulting solution has a total volume of 4560 L, containing 92 g/L of PPO, 110 mg/L of bis-MSB, and 1.58 g/L of BHT.

Following dissolution, the M.S. is pumped into an extraction tank for sequential batch extractions using acid and water. The acid extraction is performed using 2000 L of 5 wt\% nitric acid at \SI{40}{\celsius}, involving 1 hour of stirring followed by 1 hour of settling for phase separation. To minimize acid consumption, the nitric acid is recovered and reused for up to 10 batches before being discarded. Subsequently, the M.S. undergoes a two-stage water extraction using HPW, primarily to eliminate residual nitric acid. In each stage, 4000 L of HPW is added, stirred for 30 minutes, and allowed to settle for 30 minutes before the aqueous phase is drained.

After the extraction process, the absorption spectrum of the M.S. is measured. Only batches that meet the stringent optical specifications are transferred to the M.S. product tank for LS production. From there, the M.S. is pumped at a flow rate of 190 L/h through a two-stage filtration system, comprising a 200 nm filter and a subsequent 50 nm filter. The purified M.S. is then diluted with LAB to formulate the standard LS. Finally, the standard LS is delivered through a 1300 m pipeline at a flow rate of 7000 L/h to the underground LS hall for further processing.

\section{Composition and Design Requirements of the Mixing Plant}\label{sec3}

\begin{figure}[h]
\centering
\includegraphics[width=1\textwidth]{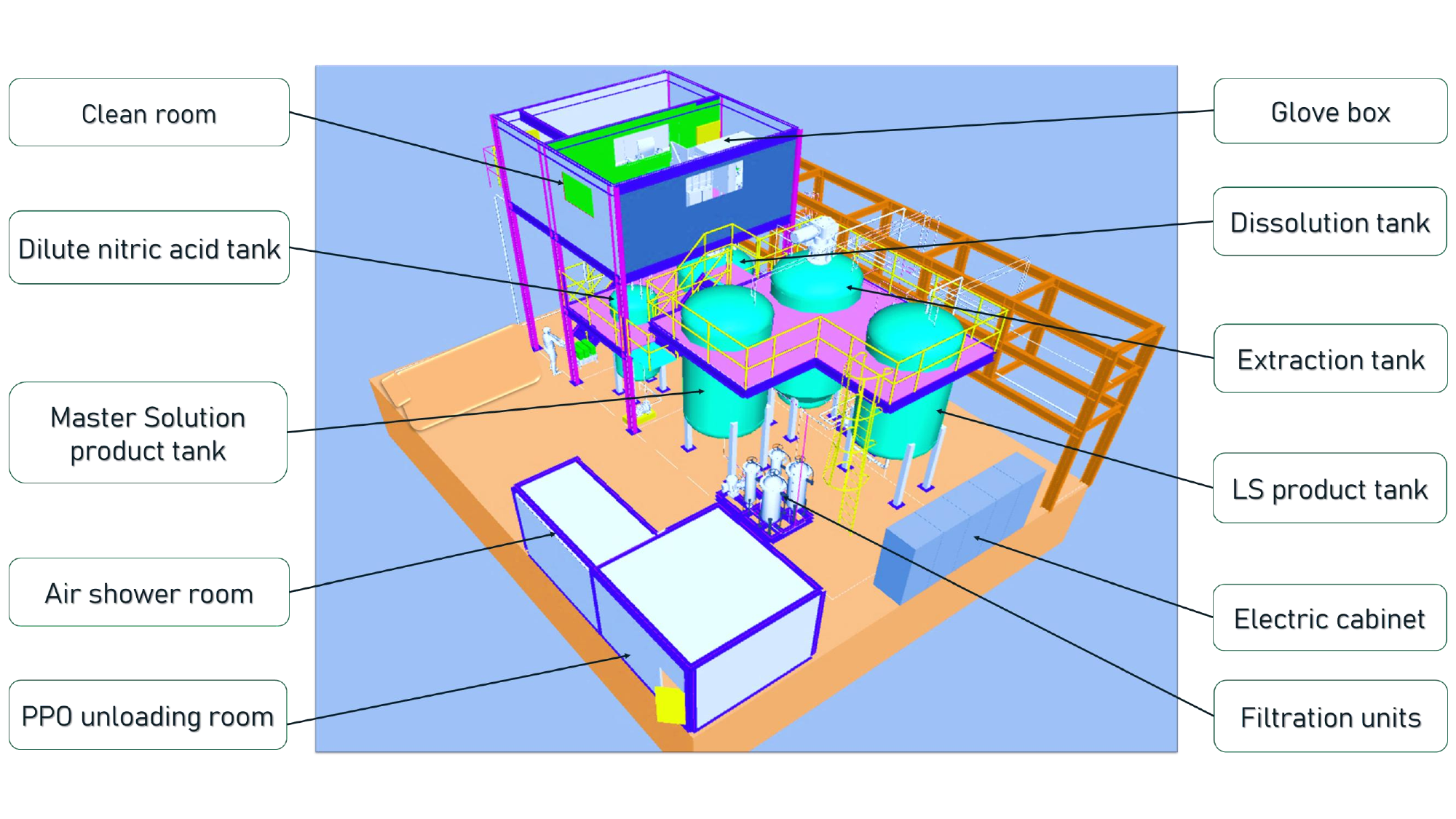}
\caption{Spatial Configuration of the Mixing Plant. A 3D architectural rendering identifying the key functional zones: the Clean Room, the Glove Box for solute feeding, the Dissolution and Extraction Tanks, and the filtration units.}\label{fig3}
\end{figure}

The conceptual design of the Mixing Plant is illustrated in Fig.~\ref{fig3}. The main components of the plant are a feeding glove box, a clean room, a dissolution tank, an extraction tank, a M.S. product tank, filtration units, an LS product tank, a dilute nitric acid tank, a PPO unloading room, an air shower room, and an electrical cabinet.

Auxiliary components consist of stainless steel pipelines, centrifugal pumps, diaphragm pumps, vacuum pumps, valves, flow meters, level sensors, thermometers, hot water units, a nitrogen sealing module, an automatic control system based on a programmable logic controller (PLC), a PPO lifting mechanism, and waste liquid drainage pipelines. HPW, concentrated nitric acid, nitrogen gas, and network access are provided by external systems.

To ensure compatibility with LS, stainless steel (304L or 316L) is selected for tanks and pipe lines, and the internal surfaces are polished to achieve a roughness of \( \mathrm{R_a} \leq \SI{0.4}{\micro\meter} \). To reduce airborne radon contamination, the entire plant is tightly sealed with a leak rate below \( 10^{-6}~\si{\milli\bar\liter\per\second} \), and all storage tanks are continuously purged with HPN. Flange connections employ a double O-ring sealing structure, with nitrogen gas filled between the two O-rings to enhance leak protection. A nitrogen-sealed enclosure is also added around the flange as a secondary safeguard. The agitator shafts of dissolution and extraction tank is equipped with a magnetic fluid seal, in which the fluid used is compatible with the LS. A monitoring device is also integrated to provide real-time sealing performance feedback.

The glove box is an essential device for feeding solutes into the system, connected to the dissolution tank through a pipe with a diameter of 20 cm. The feeding pipe is equipped with two isolation valves, whose opening must simultaneously satisfy two conditions: the pressure inside the dissolution tank must equal that inside the glove box, and the oxygen concentration within the glove box must be below 2000 ppm (corresponding to less than 1\% residual air). When these conditions are satisfied, contamination of the LS from radon and dust can be effectively minimized. During the commissioning of the Mixing Plant, the oxygen concentration in the glove box was controlled below 100 ppm, and surpassed the design requirement by one order of magnitude.

The glove box consists of a main chamber and a transfer chamber. The main chamber is connected to the dissolution tank. The transfer chamber of the glove box is equipped with an inner and an outer gate, and at least one of these two gates must remain closed; otherwise, the glove box, and even the tank, would be directly connected to the outside environment, leading to contamination. The solute is first introduced into the transfer chamber through the outer gate, while the inner gate of the transfer chamber remains closed. In the transfer chamber, the solute undergoes four nitrogen circulation purging cycles. In each cycle, the pressure is pumped down from standard atmospheric pressure to below 1000 Pa and then restored to atmospheric pressure. After the purging is completed, the inner gate is opened and the solute is transferred into the main chamber for feeding. At this time, the outer gate remains closed, so that ambient air cannot enter the main chamber or the dissolution tank.

The automatic control system, based on a PLC, is employed to control the Mixing Plant. Each tank is equipped with level, temperature, and pressure sensors. The PLC automatically regulates the liquid level, temperature, and pressure in the tanks according to real-time sensor data. In addition, flow meters and control valves are installed on the LAB and M.S. pipelines. The valve openings are adjusted by the PLC based on the feedback of the flowmeters, enabling automatic regulation of pipeline flow rates.

The dissolution tank and extraction tank are both heated by a water bath, each equipped with an independent electric heating module. Considering that the M.S. is prone to aging at high temperatures, the temperature during dissolution and extraction is maintained at approximately \SI{40}{\celsius}. Additionally, an insulation layer is installed around the M.S. product tank to prevent crystallization and pipeline blockage caused by low ambient temperatures during winter.

The extraction tank is equipped with a level sensor containing two floats to measure the total liquid level and the organic/aqueous phase interface, respectively. The height of the interface is an important parameter for evaluating the mixing performance during the extraction process. The agitator is configured with a three-stage impeller and an adjustable rotation speed. Commissioning tests indicated that, at 80\% of the rated speed(about 72 rpm), the aqueous phase (dilute nitric acid/water) and the organic phase (M.S.) were fully mixed, and no distinct phase boundary was observed on the level sensor.

After completion of the extraction process, the discharge of waste acid and wastewater is carried out through a combination of automated coarse drainage and manual fine drainage. In the coarse drainage stage, a large-diameter pipeline is used to achieve a high flow rate, with the PLC system automatically stopping the drainage based on the liquid level parameters. Subsequently, fine drainage is performed manually through a small-diameter pipeline at a low flow rate, with the organic/aqueous phase interface monitored via an observation window. The loss of M.S. during each drainage operation is controlled to less than 0.5 L.

The strict cleanliness of the equipment is required for the production of LS. Accordingly, rigorous cleaning procedures and quality inspection standards are established. All stainless-steel surfaces in contact with the LS, including tanks, pipelines, pumps, and valves, are subjected to degreasing and precision cleaning. Each tank and the entire plant are equipped with a self-circulation loop to facilitate in-situ cleaning. To meet these cleanliness specifications, the selection of pumps, valves, and other components is tightly controlled, and specific requirements are defined for the welding processes of pipelines and tanks.

\section{Construction and Installation of the Mixing Plant}\label{sec4}




After the design of the Mixing Plant was completed and passed the pre-production review, the project entered the construction phase. We deployed on-site personnel to monitor the progress and quality of the construction, focusing on the welding quality of the tanks and pipelines, as well as the surface roughness of the tank interiors.

Before stainless steel materials can be used, they must be measured for radioactive content using a Low-background high-purity germanium(HPGe) spectrometer \cite{cite_69}. Only materials with a $^{238}\text{U}$ content below 10 ppb\cite{cite_68} are deemed acceptable for use. The \(^{238}\text{U}\) concentration of stainless steel materials for the Mixing Plant is shown in Table \ref{tab4}.

\begin{table}[h]
\centering
\caption{\(^{238}\text{U}\) concentration measurement results of stainless steel materials for the Mixing Plant}
\label{tab4}
\renewcommand{\arraystretch}{1.4} 
\begin{tabular}{|c|c|}
\hline
\diagbox{materials}{Measured Parameters} & \(^{238}\text{U}\) (ppb) \\
\hline
Stainless steel plate & $< 6.4$  \\
\hline
Stainless steel piping & $< 5.6$ \\
\hline
Welding wire & $< 5.6$  \\
\hline
\end{tabular}
\end{table}

Lanthanum tungsten welding electrodes were selected, while thorium tungsten electrodes were prohibited. We use a vacuum cleaner to promptly remove the slag generated during welding, in order to prevent these particles from scratching the surface of the tank. After welding, we use mechanical polishing and electrolytic polishing to reduce the roughness, which can prevent radon from adsorbing on the inner surface of the equipment.


For the pipeline welding, argon gas shielded automatic welding technology without filler material was used. The acceptable weld profiles for groove welds on metallic tube-to-tube butt joints must comply with the ASME BPE-2016 standard\cite{ASME_BPE_2016}. The weld color should not exceed level 3 of weld discoloration on the inside of the austenitic stainless steel tube. The pipelines underwent the vacuum helium leak test, while the tanks, which could not be evacuated, underwent the positive pressure helium leak test with a suction gun.

Once all pipelines and tanks were completed, a full-scale assembly and functional testing were performed at the factory. The final step was cleaning: each tank was individually cleaned and inspected, and after passing the inspection, it was dried and sealed with nitrogen under positive pressure for storage. The pipelines were cleaned sequentially, and each cleaned pipeline was examined using an endoscope. After passing the examination, the pipelines were dried and all interfaces were sealed with blind flanges for storage.

After the construction was completed and passed manufacturer's construction acceptance review, all equipment was transported to JUNO site for installation. During the installation process, the overground LS hall was thoroughly cleaned and equipped with devices such as vacuum cleaners and Fan Filter Units. Before being moved into the hall, all external surfaces of the equipment were washed using high-pressure water. At the end of each workday, the floor was wiped down, and the cleanliness of the air was measured before work started the next day. The measurement results showed that the cleanliness level of the hall reached class 300,000 during installation. Additionally, all installation workers were trained in clean installation practices to avoid contamination of the pipes and tanks. Workers were required to wear clean gloves when connecting flanges, full-body cleanroom suits when entering tanks, and promptly remove any contamination if noticed.

After installation, five rounds of helium leak testing were performed. The number of leakage points gradually decreased, and in the end all measured leak rates met the required standard. To avoid flange leakage caused by vibration, all pipelines were reinforced. A temperature-related leak was observed at the joint between the stainless-steel pipeline and the pneumatic diaphragm pump at the outlet of the M.S. product tank. This was due to the large difference in thermal expansion between the polytetrafluoroethylene (PTFE) pump body and stainless steel. By adding a flexible fluororubber O-ring at the connection, the problem was resolved and the leak rate was reduced to within the acceptable limit.

\section{Commissioning and results of the Mixing Plant}\label{sec5}

After leak-tightness was verified, self-commissioning with HPW was performed. And the class 50 requirement of the MIL-STD 1246C, which was adopted in JUNO as cleanliness requirement for the surfaces in contact with the LS. 
Each tank was flushed in sequence, and the particle counts in the water discharged from each tank were measured to assess cleanliness. Since cleanliness was strictly controlled during installation, the self-commissioning with HPW proceeded smoothly. After each tank had been flushed three times, all particle counts met the required limit. Meanwhile, pumps, agitators, flowmeters, and sensors performed as expected, and all system functions met the design requirements.

Following the HPW stage, each tank underwent self-commissioning with LAB. During this process, samples were taken for absorption spectroscopy. A tank was considered successfully commissioned if the absorption spectra of the LAB at the inlet and outlet overlapped. All tanks passed this test in a single round of LAB self-commissioning. After self-commissioning with both HPW and LAB was completed, the system was operated together with the other subsystems to begin joint commissioning.

 During the joint commissioning in July 2024. we measured the removal efficiency of \(^{238}\text{U}\) and \(^{232}\text{Th}\) from the M.S. by filtration using the Inductively coupled plasma-Mass Spectrometry(ICP-MS)\cite{LI2025112579}. In the experiment, a two-stage setup, consisting of a 200 nm filter followed by a 50 nm filter, was used to filter the raw M.S. directly. The 200 nm filter was employed to remove large particles and to prevent the rapid blockage of the 50 nm filter, while the 50 nm filter was applied to retain smaller particles and served as the final step in the purification process. Samples of the filtrate were collected and analyzed. The results, as shown in Table ~\ref{tab2}, indicated that when the initial concentrations of \(^{238}\text{U}\) and \(^{232}\text{Th}\) were at the ppq (\num{1e-15} g/g) level, they could be further reduced by more than one order of magnitude after the two-stage filtration. Based on this result, the two-stage filtration method was adopted in the final production of LS.

\begin{table}[h]
\centering
\caption{Comparison of \(^{238}\text{U}\) and \(^{232}\text{Th}\) Concentrations Before and After Filtration}
\label{tab2}
\renewcommand{\arraystretch}{1.4} 
\begin{tabular}{|c|c|c|}
\hline
\diagbox{Samples}{Measured Parameters} & \(^{238}\text{U}\) (ppq) & \(^{232}\text{Th}\) (ppq) \\
\hline
Raw M.S. & \num{7.8 \pm 0.8} & \num{4.4 \pm 0.5} \\
\hline
M.S. after Filtration & \num{0.50 \pm 0.06} & \( <0.25 \)\\
\hline
\end{tabular}
\end{table}

After a series of commissioning, we optimized both the production parameters and the PLC control logic. This allowed the system to produce LS in a stable and efficient manner.

Table ~\ref{tab3} summarizes the measured concentrations of \(^{238}\text{U}\) and \(^{232}\text{Th}\), as well as the attenuation length (A.L.), from the final round of joint commissioning in November 2024. The results indicated that when the initial concentrations of \(^{238}\text{U}\) and \(^{232}\text{Th}\) in the M.S. were at the ppq  level, they could be further reduced by approximately one order of magnitude through acid plus water extraction. And the LS has met the requirements of the JUNO experiment, indicating that the Mixing Plant is ready for formal LS production.

\begin{table}[h]
\centering
\caption{Summary of \(^{238}\text{U}\) and \(^{232}\text{Th}\) Concentrations and Attenuation Length}
\label{tab3}
\renewcommand{\arraystretch}{1.4} 
\begin{tabular}{|c|c|c|c|}
\hline
\diagbox{Samples}{Measured Parameters} & \(^{238}\text{U}\) (ppq) & \(^{232}\text{Th}\) (ppq) & A.L. (m) \\
\hline
Raw M.S. & \num{2.6 \pm 0.3} & \num{4.5 \pm 0.5} & N/A \\
\hline
M.S. after Acid plus Water Extraction & \num{0.15 \pm 0.02} & \num{0.57 \pm 0.06} & N/A \\
\hline
M.S. after Filtration & \num{0.14 \pm 0.02} & \num{0.16 \pm 0.02} & N/A \\
\hline
LS Product & \num{0.06 \pm 0.01} & \num{0.05 \pm 0.01} & \num{20.33 \pm 1.09} \\
\hline
\end{tabular}
\end{table}

\section{Conclusion}\label{sec6}
The Mixing Plant, which is designed to accurately mix solutes in defined proportions and dissolve them into LAB, is one of the core components in the production of ultrapure LS. In the purification process of the Mixing Plant, extraction and filtration are employed to reduce the concentrations of \(^{238}\text{U}\) and \(^{232}\text{Th}\) in the M.S. by approximately two orders of magnitude.

Following the phases of process development, system design and construction, and installation and commissioning, the Mixing Plant has been validated for its ability to produce LS in compliance with the stringent requirements of the JUNO experiment. LS production has been ongoing since February 2025. Further details on the production process will be presented in subsequent publications.


\bmhead{Acknowledgements}
This research is supported by Strategic Priority Research Program of the Chinese Academy of Sciences (XDA10010500).

\bmhead{Declaration of competing interest}
The authors declare that they have no known competing financial interests or personal relationships that could have appeared to influence the work reported in this paper.

\bibliographystyle{sn-mathphys-num}
\bibliography{sn-bibliography}

@article{JUNO:2015sjr,
    author = "Djurcic, Zelimir and others",
    collaboration = "JUNO",
    title = "{JUNO Conceptual Design Report}",
    eprint = "1508.07166",
    archivePrefix = "arXiv",
    primaryClass = "physics.ins-det",
    month = "8",
    year = "2015"
}

@article{He_2023,
doi = {10.1088/1748-0221/18/02/P02013},
url = {https://dx.doi.org/10.1088/1748-0221/18/02/P02013},
year = {2023},
month = {feb},
publisher = {IOP Publishing},
volume = {18},
number = {02},
pages = {P02013},
author = {He, Miao and Qin, Zhonghua and Hou, Shaojing and Jing, Xiaoping and Liu, Hongbang and Ke, Zunjian and Wu, Diru and Xie, Wan and Xu, Meihang and Chen, Fang and Lu, Junguang and Heng, Yuekun and Zhang, Jiawen and Ma, Xiaoyan and Du, Zhipeng},
title = {Design of the PMT underwater cascade implosion protection system for JUNO},
journal = {Journal of Instrumentation}
}

@article{JUNO:2021kxb,
    author = "Abusleme, Angel and others",
    collaboration = "JUNO",
    title = "{Radioactivity control strategy for the JUNO detector}",
    eprint = "2107.03669",
    archivePrefix = "arXiv",
    primaryClass = "physics.ins-det",
    doi = "10.1007/JHEP11(2021)102",
    journal = "JHEP",
    volume = "11",
    pages = "102",
    year = "2021"
}

@article{JUNO_PPNP,
title = {JUNO physics and detector},
journal = {Progress in Particle and Nuclear Physics},
volume = {123},
pages = {103927},
year = {2022},
issn = {0146-6410},
doi = {https://doi.org/10.1016/j.ppnp.2021.103927},
url = {https://www.sciencedirect.com/science/article/pii/S0146641021000880}
}

@article{LI2025112579,
title = {A practical approach of measuring $^{238}\text{U}$ and $^{232}\text{Th}$ in liquid scintillator to sub-ppq level using ICP-MS},
journal = {Radiation Physics and Chemistry},
volume = {230},
pages = {112579},
year = {2025},
issn = {0969-806X},
doi = {https://doi.org/10.1016/j.radphyschem.2025.112579},
url = {https://www.sciencedirect.com/science/article/pii/S0969806X25000714},
author = {Yuanxia Li and Jie Zhao and Yayun Ding and Tao Hu and Jiaxuan Ye and Jian Fang and Liangjian Wen}
}

@article{ZHU2023167890,
title = {Optical purification pilot plant for JUNO liquid scintillator},
journal = {Nuclear Instruments and Methods in Physics Research Section A: Accelerators, Spectrometers, Detectors and Associated Equipment},
volume = {1048},
pages = {167890},
year = {2023},
issn = {0168-9002},
doi = {https://doi.org/10.1016/j.nima.2022.167890},
url = {https://www.sciencedirect.com/science/article/pii/S0168900222011822},
author = {Zhihang Zhu and Lijun Sun and Tao Hu and Jian Fang and Jiaxuan Ye and Li Zhou and Mengchao Liu and Wanjin Liu and Xiao Cai and Xilei Sun and Yayun Ding and Yuguang Xie and Boxiang Yu}
}

@article{LANDINI2024169887,
title = {Distillation and gas stripping purification plants for the JUNO liquid scintillator},
journal = {Nuclear Instruments and Methods in Physics Research Section A: Accelerators, Spectrometers, Detectors and Associated Equipment},
volume = {1069},
pages = {169887},
year = {2024},
issn = {0168-9002},
doi = {https://doi.org/10.1016/j.nima.2024.169887},
url = {https://www.sciencedirect.com/science/article/pii/S0168900224008131},
author = {C. Landini and M. Beretta and P. Lombardi and others}
}

@article{YE2022166251,
title = {Development of water extraction system for liquid scintillator purification of JUNO},
journal = {Nuclear Instruments and Methods in Physics Research Section A: Accelerators, Spectrometers, Detectors and Associated Equipment},
volume = {1027},
pages = {166251},
year = {2022},
issn = {0168-9002},
doi = {https://doi.org/10.1016/j.nima.2021.166251},
url = {https://www.sciencedirect.com/science/article/pii/S016890022101086X},
author = {Jiaxuan Ye and Jian Fang and Li Zhou and Wei Hu and Wanjin Liu and Yayun Ding and Meangchao Liu and Boxiang Yu and Xilei Sun and Lijun Sun and Yuguang Xie and Xiao Cai and Zhihang Zhu and Tao Hu}
}

@article{LING2024111305,
title = {JUNO high purity nitrogen plant},
journal = {Applied Radiation and Isotopes},
volume = {208},
pages = {111305},
year = {2024},
issn = {0969-8043},
doi = {https://doi.org/10.1016/j.apradiso.2024.111305},
url = {https://www.sciencedirect.com/science/article/pii/S0969804324001337},
author = {Xin Ling and Boxiang Yu and Zhilong Hou and Tao Hu and Li Zhou and Yongjun Yan and Zhihang Zhu and Xiao Cai and Yayun Ding and Jian Fang and Junyu shao and Lijun Sun and Xilei Sun and Yuguang Xie and Xiaohui Qi and Haodong Zhang}
}

@article{cite_69,
doi = {10.1088/1674-1137/39/8/086002},
url = {https://doi.org/10.1088/1674-1137/39/8/086002},
year = {2015},
month = {aug},
publisher = {Chinese Physical Society and the Institute of High Energy Physics of the Chinese Academy of Sciences and the Institute of Modern Physics of the Chinese Academy of Sciences and IOP Publishing},
volume = {39},
number = {8},
pages = {086002},
author = {Niu, Shun-Li and Cai, Xiao and Wu, Zhen-Zhong and Liu, Yi and Xie, Yu-Guang and Yu, Bo-Xiang and Wang, Zhi-Gang and Fang, Jian and Sun, Xi-Lei and Sun, Li-Jun and Liu, Ying-Biao and Gao, Long and Zhang, Xuan and Zhao, Hang and Zhou, Li and Lü, Jun-Guang and Hu, Tao},
title = {Simulation of background reduction and Compton suppression in a low-background HPGe spectrometer at a surface laboratory},
journal = {Chinese Physics C}
}

@techreport{cite_68,
title = {Technote of the JUNO cleanliness control programs},
author = {Jie Zhao and Gaosong Li and others}
}

@book{ASME_BPE_2016,
  author       = {The American Society of Mechanical Engineers},
  title        = {ASME BPE-2016: Bioprocessing Equipment},
  year         = {2016},
  pages        = {220},
  url          = {https://studylib.net/doc/26101437/toaz.info-asme-bpe-2016}
}

\end{document}